\documentclass{aa}  

\usepackage{graphicx}
\usepackage{txfonts}

\usepackage{booktabs}

\usepackage{multicol}
\usepackage{subfig}
\usepackage{hyperref}
\usepackage{gensymb}
\usepackage{xcolor}

\begin{document}

\title{The chemo-dynamical complexity of $\omega$ Centauri: \\ Different kinematics for different populations}
  \titlerunning{The chemo-dynamical complexity of $\omega$ Cen}

   \author{G.~Pagnini
          \inst{1, 2}
          \and
          P. ~Bianchini\inst{1}
         \and
          P.~Di Matteo\inst{3}
          \and
          A.~Mastrobuono-Battisti\inst{4,5,6}
        }

   \institute{Université de Strasbourg, CNRS, Observatoire Astronomique de Strasbourg, UMR 7550, F-67000 Strasbourg, France\\ \email{pagnini@astro.unistra.fr}
   \and Centre national d’études spatiales (CNES), 2, Place Maurice Quentin, 75039, Paris, France
   \and LIRA, Observatoire de Paris, Université PSL, Sorbonne Université, Université Paris Cité, CY Cergy Paris Université, CNRS, 92190 Meudon, France
\and Dipartimento di Fisica e Astronomia “Galileo Galilei”, Università di Padova, Vicolo dell’Osservatorio 3, 35122 Padova, Italy
        \and Dipartimento di Tecnica e Gestione dei Sistemi Industriali, Università degli Studi di Padova, Stradella S. Nicola 3, I-36100 \\Vicenza, Italy
        \and Istituto Nazionale di Astrofisica – Osservatorio Astronomico di Padova, Vicolo dell’Osservatorio 5, Padova, I-35122, Italy
}
   \date{Received xxx; accepted xxx}

  \abstract
  {The origin of $\omega$~Centauri remains one of the key open questions in
stellar dynamics and chemical evolution. Its large abundance spreads
and multiple populations suggest a formation history more complex than
that of a typical globular cluster.}{We investigate whether the chemical sub-populations identified in
APOGEE DR17 also exhibit distinct spatial and kinematic signatures and
the constraints these provide on the formation pathways of
$\omega$~Cen.}{We analysed a sample of APOGEE DR17 red giant stars using a Gaussian mixture model in an 8D chemical-abundance space. The resulting chemical components were combined with Gaia proper motions and APOGEE line-of-sight velocities to derive intrinsic mean velocities and velocity dispersions in all three observable directions. We measured both global kinematic quantities and radial profiles for each chemically defined group, extending from the inner regions to $\sim4$ half-light radii. }{The Gaussian mixture model identifies five chemical components, which, when examined through their radial cumulative distributions, naturally group into two broader families characterised by lower and higher
aluminium enrichment. The two families differ significantly in their spatial and kinematic properties: the Al-rich stars are more centrally concentrated and exhibit stronger radial anisotropy than the Al-poor stars, which remain closer to isotropy over the radial range probed. Despite these significant differences (> 2$\sigma$), the two populations share a common rotation pattern, with comparable intrinsic rotation amplitudes and rotation-axis inclinations within the uncertainties ($\sim$~0.5 km\,s$^{-1}$).  Both families span a wide range in metallicity, indicating that they do not simply correspond to a first- and second-generation dichotomy but rather trace chemically and dynamically complex substructures within $\omega$~Cen.}{This work represents the first chemo-dynamical study of $\omega$~Cen linking detailed chemical tagging to internal kinematics from the inner regions to the cluster outskirts and provides a key benchmark for models of its formation. A formation path involving both hierarchical assembly within a dwarf-galaxy potential and centrally concentrated, chemically enriched star formation offers a natural explanation for the observed chemo-dynamical complexity. Extending this approach to larger fields of view and to more numerous, fainter stars will be essential to robustly uncover the origin of this uniquely complex system.}

   \keywords{globular clusters: individual (NGC 5139) - stars: abundances - stars: kinematics and dynamics
               }

   \maketitle
%

\section{Introduction}

Omega Centauri (\(\omega\)~Cen, NGC~5139) is the most massive---and arguably the most puzzling---globular cluster (GC) in the Milky Way, with a present-day mass of \(M \simeq 3.55 \times 10^{6}\,M_\odot\) \citep[][]{baumgardt18}. Long before the modern view of the ubiquitous phenomena of multiple stellar populations (MPs) in GCs, \(\omega\)~Cen was known for star-to-star abundance variations (e.g. \citealt{norris75,butler78,persson80}) and for photometric splits along the red giant branch (RGB), main sequence (MS), and subgiant branch (SGB; e.g. \citealt{Lloydevans1977,Bedin2004,lee99}). Spectroscopy later revealed a wide internal metallicity range---from \([\mathrm{Fe/H}] \approx -2\)  to \([\mathrm{Fe/H}] \gtrsim -0.7\)---together with increasing s-process elements and increasing total C+N+O at higher \([\mathrm{Fe/H}]\) \citep[e.g.][]{norris95,johnson2010,marino11,marino12,nitschai23}. At fixed metallicity, \(\omega\)~Cen shows the familiar GC light-element (anti-)correlations (C--N, Na--O, Mg--Al) and helium variations; the strongest effects often occur at intermediate \([\mathrm{Fe/H}]\) \citep[][]{johnson2010,marino11,marino12,clontz2025}. The reported age spreads range from \(\sim\!2\)~Gyr to \(\lesssim 0.5\)~Gyr, with C+N+O variations complicating SGB-based estimates \citep[][]{hilker04,villanova07,villanova14, tailo2016, clontz2024, wang2026}. All of these chemical and age peculiarities point to an origin in a substantially more massive parent system, with $\omega$~Cen most naturally interpreted as the bound nuclear remnant of a disrupted dwarf galaxy rather than a typical Galactic GC \citep{lee99,majewski00,carraro00,bekki03,tsuchiya03,tsuchiya04, souza2026}.

Colour--magnitude diagrams (CMDs) of \(\omega\)~Cen reveal a remarkably rich set of sequences at all evolutionary stages
\citep[e.g.][]{pancino00,Bedin2004,ferraro2004,bellini2010,bellini2017b,milone2017,clontz2025}. To disentangle these features at a fixed luminosity, \citet{milone2015} introduced the `chromosome map' (ChM), a pseudo two-colour diagram built from HST F275W, F336W, F438W, and F814W photometry. After `verticalising' the underlying CMDs, stars are placed in the plane \((\Delta C_{\mathrm{F275W,F336W,F438W}},\,\Delta(\mathrm{F275W{-}F814W}))\), where the first coordinate is primarily sensitive to N and other light-element anomalies, and the second traces He and overall metallicity. In this space, metal-poor RGB stars define a compact first-population (1P) clump and an extended second-population (2P) sequence towards higher \(\Delta C_{\mathrm{F275W,F336W,F438W}}\).  At higher \([\mathrm{Fe/H}]\), both 1P and 2P split into a `lower stream' (LS) and an `upper stream' (US) that extend to large \(\Delta(\mathrm{F275W{-}F814W})\); LS stars retain 1P-like light elements at fixed metallicity, whereas US stars display 2P-like patterns \citep[][]{milone2017b,marino2019,clontz2025}.

While the ChM is sensitive to light-element and He variations, it is not a direct abundance measurement; spectroscopy is therefore essential to anchor the photometric tags to element-by-element trends.
High-resolution spectroscopic surveys have mapped element-by-element trends across \(\omega\)~Cen’s [Fe/H] range, refining the light- and heavy-element patterns and highlighting sub-populations \citep[e.g.][]{johnson2010,marino11,marino12}. Large spectroscopic surveys such as APOGEE have further expanded this view by providing homogeneous abundance measurements for large stellar samples. For example, \citet{meszaros2021} analysed APOGEE DR16 spectra of $\omega$~Cen using the BACCHUS pipeline and confirm the presence of multiple chemically distinct stellar populations spanning a wide metallicity range. More recently, \citet{Mason2025} combined APOGEE DR17 abundances with \textit{Gaia}, MUSE, and HST data to chemically tag \(\omega\)~Cen stars. They identify three chemically distinct groups—(i) a population resembling metal-poor halo stars, (ii) an intermediate group with narrow \([\mathrm{Fe/H}]\) and classic GC light-element anti-correlations, and (iii) an extreme second-generation population. They show that these alone reproduce the observed ChM distribution. This result links discrete ChM features to chemically defined groups, showing that a small set of abundance patterns can reproduce the observed photometric substructure.

Using high-resolution optical spectroscopy, \citet{alvarezgaray2024} further investigated the Mg--Al abundance patterns of $\omega$~Cen giants and show that Al-enhanced populations are preferentially concentrated towards the cluster centre. Similarly, by combining complementary photometric and spectroscopic datasets, \citet{dondoglio2025} identify chemically distinct stellar groups across $\omega$~Cen from the centre out to $\sim30'$ ($\approx5\,r_h$), showing that these populations are not fully spatially mixed but instead display clear radial gradients. In particular, stars enriched in light elements are found to be more centrally concentrated than their chemical counterparts. If these chemical and spatial differences are the results of \(\omega\)~Cen's complex formation history, one might also expect differences in the stellar kinematics and more generally in chemo-dynamical space.

The kinematic picture of \(\omega\)~Cen has progressed from small bright-giant samples to panoramic phase-space maps. Early work used line-of-sight velocities for a few hundred giants \citep{Suntzeff1996,Mayor1997,Reijns2006} and ground-based proper motions for bright stars \citep{vanLeeuwen2000}. Multi-epoch HST data then provided proper motions for hundreds of thousands of stars down to faint MS levels \citep{anderson2010,Bellini2014,Bellini2018,haberle2025}, and MUSE added spectra for comparably large samples \citep{Bacon2010,Kamann2018,nitschai23,pechetti24}. Beyond the half-light radius, US stars show more radial anisotropy than LS stars \citep{cordoni2020}. At \(\sim\!3.5\,r_h\), 2P stars are more radially anisotropic than 1P \citep{Bellini2018}. These contrasts weaken when stars are sorted only by \([\mathrm{Fe/H}]\) \citep{cordoni2020}, and inside the half-light radius stars with different iron abundance can share similar kinematics \citep{vernekar2025, haberle2025}. Collectively, these results indicate that kinematic differences are more tightly coupled to light-element population identity than to iron abundance alone, at least over the radii probed so far.
 
In the context of nuclear star cluster (NSC) formation (e.g. \citealt{neumayer2020}), the main mechanisms commonly invoked include hierarchical assembly through the inspiral and merging of stellar systems, centrally concentrated star formation sustained by gas retention and chemical enrichment, and hybrid `wet merger' scenarios in which cluster merging is accompanied by gas inflow and subsequent star formation (e.g. \citealt{guillard2016}).

Recent advances in large spectroscopic and astrometric surveys now offer a new opportunity to investigate the formation of $\omega$~Cen through a fully multi-dimensional approach. High-resolution chemical abundances from APOGEE, combined with precise \emph{Gaia} proper motions and line-of-sight velocities from APOGEE, make it possible to jointly characterise stellar populations in chemical-abundance space and in 3D kinematics over a wide field of view. This combination provides the completeness and precision required to explore how chemical enrichment, spatial structure, and orbital properties are coupled within the cluster from the inner regions to the outskirts. In this work, we exploit these new data to perform a chemo-dynamical analysis of $\omega$~Cen, with the aim of constraining its formation path and discriminating between scenarios dominated by mergers of GCs, centrally concentrated in situ chemical enrichment, or a combination of both.

This paper is structured as follows. In Sect.~\ref{obsdata}, we discuss the observational sample analysed in this work. In Sect.~\ref{analysis} we explain the chemo-dynamical analysis, and in Sect.~\ref{results} we present the results. In Sect.~\ref{discussion}, we discuss the novelties of our results also in the context of other recent studies on the subject. Finally, in Sect.~\ref{conclusions} we derive our conclusions.


\section{Observational data}\label{obsdata}

For this study, we made use of data from the APOGEE Value Added Catalogue (VAC) of Galactic GC stars \citep{schiavon2023}.  From this catalogue, we selected stars associated with $\omega$~Centauri by applying the following criteria:
\begin{enumerate}
\item a high membership probability according to \citet{vasiliev21}
      (\texttt{VB\_PROB} $\geq 0.9$);
\item a signal-to-noise ratio $\texttt{SNREV} > 70$;
\item effective temperatures in the range
      $3500\,\mathrm{K} < T_{\mathrm{eff}} < 5500\,\mathrm{K}$ and surface
      gravities $\log g < 3.6$;
\item APOGEE quality cuts requiring \texttt{APOGEE\_STARFLAG} = 0 and \texttt{APOGEE\_STARBAD} = 0.
\end{enumerate}
This defines a parent sample of 1201 stars. For each of them, the APOGEE VAC provides detailed chemical abundances, line-of-sight velocities, and \emph{Gaia}~EDR3 proper motions \citep{gaiaEDR3}. We subsequently retained only stars with valid abundance measurements for all eight elements simultaneously, namely Fe, Mg, Si, Ca, C, Al, K, and Mn, which were used in the Gaussian mixture analysis described below. The final sample used in this work therefore consists of 588 red giant stars, spanning projected clustercentric distances out to $\sim 40$ arcmin, thus covering a large fraction of the cluster field of view and extending well up to $\sim 8\,r_h$.
  
\section{Analysis}\label{analysis}
\subsection{From chemical to spatial differences}
\label{analysis_1}
\begin{figure*}\centering
 \includegraphics[clip=true, trim = 0mm 0mm 0mm 0mm, width=0.44\linewidth]{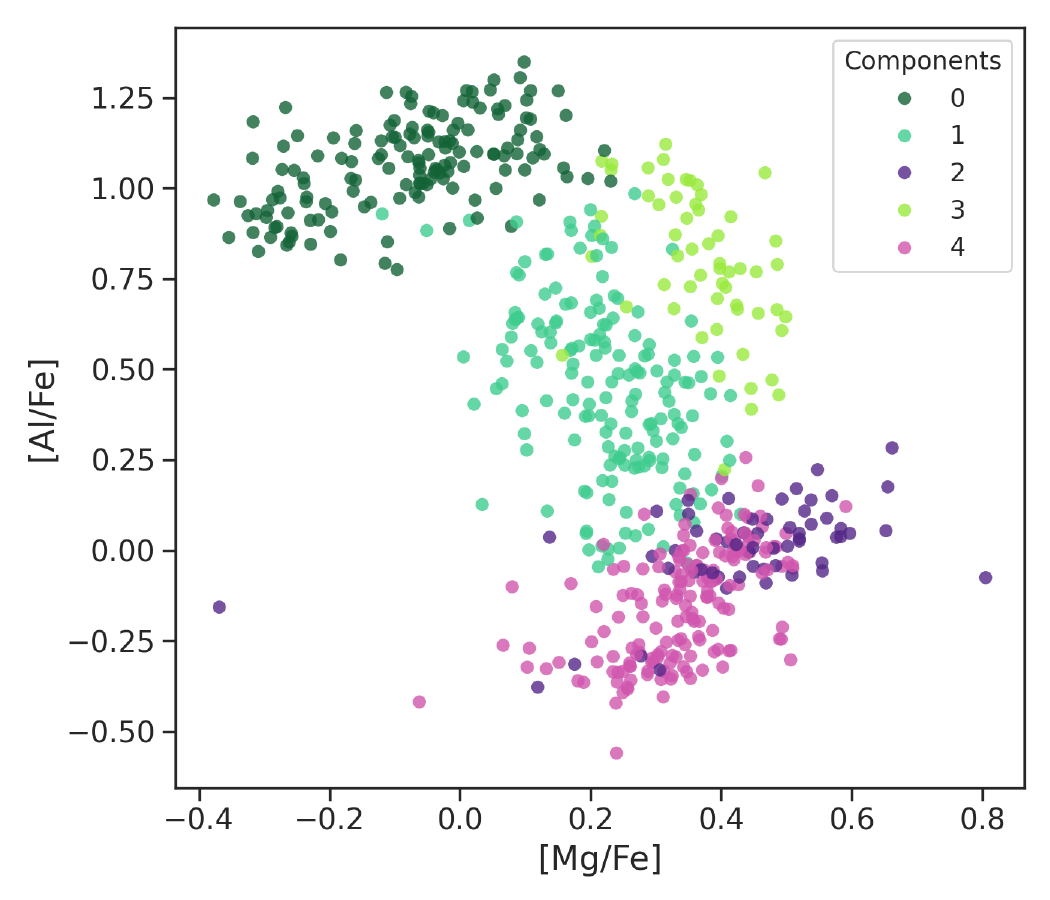}
\includegraphics[clip=true, trim = 0mm 0mm 0mm 0mm, width=0.5\linewidth]{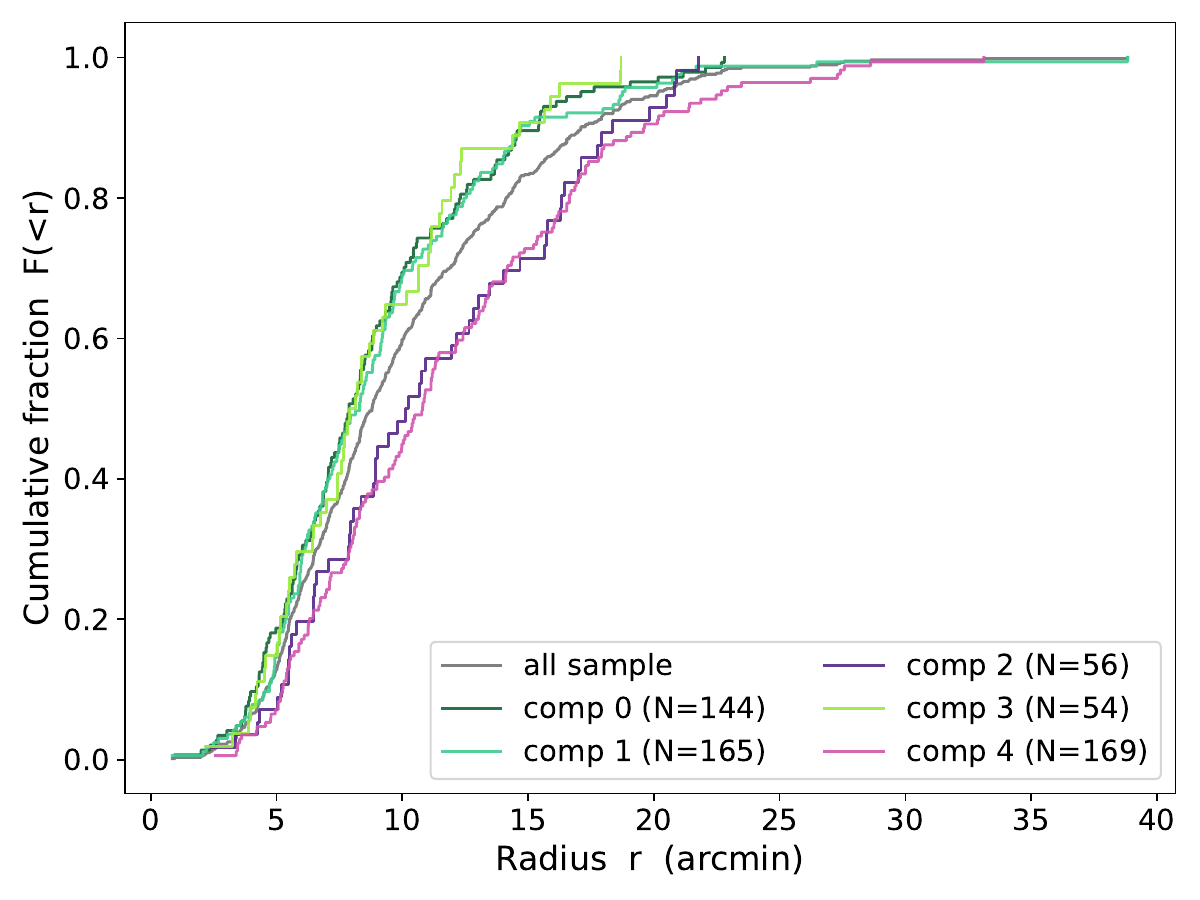}
  \caption{Chemical clustering and spatial distribution of APOGEE stars in $\omega$ Cen. Left panel: [Al/Fe] vs [Mg/Fe] distribution of stars belonging to $\omega$ Cen, colour-coded according to the different components retrieved when minimising the BIC criterion in the GMM. Right panel: sCDF of stars in the different components. For reference, the global $\omega$ Cen's CDF is shown in grey.}\label{fig:5pops}
 \end{figure*}

This work builds on the analysis of \citet{pagnini2025}, where we applied a Gaussian mixture model (GMM) approach\footnote{We used the
\texttt{GaussianMixture} class available in \texttt{scikit-learn} (\url{https://scikit-learn.org/stable/modules/generated/sklearn.mixture.GaussianMixture.html}).} to an 8D chemical-abundance space defined by [Fe/H], the $\alpha$-elements [Mg/Fe], [Si/Fe], and [Ca/Fe], the light and odd-$Z$ elements [C/Fe], [Al/Fe], and [K/Fe], and the
iron-peak element [Mn/Fe] (see Sect.~3 of \citealt{pagnini2025}).  The choice of eight elements reflects a compromise between the ability to capture chemical differences between populations and retaining a sufficiently large sample with reliable abundance measurements. 
The distribution of $\omega$ Cen in this 8D abundance space was fitted by using an increasing number of Gaussian components and the optimal number of components for our dataset was then determined by minimising the Bayesian information criterion (BIC). With this procedure, we find that the number of components that best reproduces the 8D distribution of $\omega$ Cen is five \citep[see Sect.~3 in][for discussion on how this number can vary]{pagnini2025}. To assess the impact of abundance uncertainties on this decomposition, we performed 200 Monte Carlo realisations in which the chemical abundances were perturbed according to their quoted uncertainties and the GMM analysis was repeated. In these tests, a five-component solution remains the most frequently selected one, being recovered in 129 out of 200 realisations, whereas a four-component solution is preferred in the remaining 71 cases. As already discussed in \citet{pagnini2025}, however, we caution against attributing a direct physical meaning to this exact number of components. For simplicity, we label these five GMM components as populations 0–4. Figure~\ref{fig:5pops} (left panel) shows these components in the [Al/Fe] versus [Mg/Fe] space (see Fig.~\ref{fig:8dchem} for the other chemical abundance spaces analysed in the GMM). The five GMM components occupy distinct regions of chemical-abundance space and are most clearly separated in the [Al/Fe]–[Fe/H] plane, where aluminium defines the primary axis of separation. Two components (Populations 2 and 4) are Al-poor across a broad metallicity range, consistent with first-generation-like chemistry. The remaining three components (Populations 0, 1, and 3) are Al-enhanced relative to these Al-poor groups. Population 0 traces the most chemically extreme, Al-rich sequence, while Population 1 occupies an intermediate Al-enhanced regime. Population 3 is characterised by higher metallicity combined with enhanced aluminium, forming a chemically distinct metal-rich, Al-rich component.
In this sense, our division is broadly consistent with recent independent work, such as \citet{Mason2025}: their intermediate (IM) population maps broadly onto our component 1, their P1 stars onto our components 2 and 4, and their P2 stars onto our components 0 and 3. Thus, although the exact number of groups differs, both analyses support the presence of genuine and astrophysically meaningful chemical substructure in $\omega$~Cen. We therefore used the five-component decomposition as a useful representation of the chemical-abundance space and proceeded to test whether the resulting structures also show distinct spatial and kinematic signatures.

To better understand the physical meaning behind this data driven sub-population division, we started by focusing on the spatial distribution of these five different components within $\omega$ Cen. In this regard, Fig.~\ref{fig:5pops} features the spatial cumulative distribution function (sCDF) of the five sub-populations compared to the global sCDF across the entire radial range. From this plot, we can see that these chemically identified populations converge into two distinct ones: components 0, 1, and 3 appear more spatially concentrated with respect to components 2 and 4. Interestingly, the former two also differ from the latter from a chemical point of view, as they are richer in aluminium (see left panel of Fig.~\ref{fig:5pops}). 

Motivated by these chemical and spatial differences, we decided to combine the five sub-populations into two main groups: the spatially concentrated Al-rich one with a total of 363 stars (i.e. 0+1+3 components), and the more spatially extended Al-poor one with a total of 225 stars (i.e. 2 plus 4 components). Although we refer to these two families simply as Al-poor and Al-rich, this division can be viewed, to first order, as analogous to the distinction between less chemically processed, 1P-like stars and more enriched, 2P-like stars commonly discussed in the multiple-population literature. We stress, however, that in $\omega$ Cen this correspondence is only approximate, since both families span a wide metallicity range and do not map uniquely onto the standard 1P/2P classification.

\subsection{Internal kinematics}
We analysed population kinematics by converting \emph{Gaia} proper motions into radial and tangential components around the cluster centre, after removing the bulk motion and correcting for projection effects and perspective motions across the field of view.
The procedure is the following: 

\begin{itemize}
\item We transformed equatorial coordinates $(\alpha,\delta)$ into Cartesian coordinates $(x,y)$ following Eq.~(1) of \citet{vandeven2006}, adopting $r_0 = 10800/\pi$ and the cluster centre $(\alpha_0,\delta_0)=(201.697^\circ,-47.480^\circ)$ \citep{vasiliev21}. By convention, the positive $x$-axis points west:
\begin{equation}
\begin{aligned}
x &= -\,r_0 \cos\delta \,\sin(\alpha-\alpha_0),\\
y &= \;r_0 \big[\sin\delta \cos\delta_0 - \cos\delta \sin\delta_0 \cos(\alpha-\alpha_0)\big].
\end{aligned}
\label{eq:vv06-pos}
\end{equation}

\item We converted the Gaia proper motions $(\mu_{\alpha*},\mu_\delta)$ into the same Cartesian frame using Eq. (2) of \citet{gaia18}:
\begin{equation}
\begin{aligned}
\mu_x &= -\mu_{\alpha*}\cos(\alpha-\alpha_0) + \mu_\delta \sin\delta \,\sin(\alpha-\alpha_0),\\
\mu_y  &= \mu_{\alpha*}\sin\delta_0 \,\sin(\alpha-\alpha_0)\\
       &\quad + \mu_\delta\big[\cos\delta \cos\delta_0 + \sin\delta \sin\delta_0 \cos(\alpha-\alpha_0)\big].
\end{aligned}
\label{eq:gaia-ortho}
\end{equation}
where we adopted the sign convention that proper motions are positive to the west, hence the use of $-\mu_x$ for consistency with the positional $x$ axis.

\item Before converting $(\mu_x,\mu_y)$ into radial and tangential components $(\mu_r,\mu_t)$, we subtracted the systemic proper motion of the cluster adopting values from \citet{vasiliev21} and corrected the proper motions for perspective effects arising from the bulk space motion of the cluster, following Eq.~(6) of
\citet{vandeven2006}.  Given the large angular extent and non-negligible systemic velocity of $\omega$~Cen, this correction removes the apparent solid-body rotation and velocity gradients induced purely by projection effects.

\item We then obtained centre-referenced radial and tangential proper motions using Eq.~(3) from \citet{vanLeeuwen2000}. Let $R=\sqrt{x^2+y^2}$ denote the projected distance from the cluster centre. The projections are
\begin{equation}
\begin{aligned}
\mu_r &= \frac{x\,\mu_x + y\,\mu_y}{R},\\[2pt]
\mu_t &= \frac{-\,y\,\mu_x + x\,\mu_y}{R}.
\end{aligned}
\label{eq:rt-proj}
\end{equation}
 where positive $\mu_r$ denotes motion away from the cluster centre and positive $\mu_t$ corresponds to counter-clockwise rotation. 
 The proper motions are then expressed in kilometres per second, assuming a distance of the cluster of $\rm d=5.426 \pm 0.047 \,kpc$ \citep{baumgardt21}.
\end{itemize}

For the APOGEE~DR17 line-of-sight velocities \citep{abdurro22}, we first subtracted the cluster’s systemic velocity to place the measurements in the cluster rest frame, adopting the value from \citet{vasiliev21}. We then corrected for perspective motions following Eq.~(6) of \citet{vandeven2006}. 

\subsubsection{Radial profiles}
Using the kinematic measurements described above, we computed both spatially resolved radial profiles and global kinematic quantities, with the goal of identifying chemo-dynamical differences between the Al-poor and Al-rich populations. We measured the internal kinematics of $\omega$~Cen by deriving the mean velocities $v$ and velocity dispersions $\sigma$ in the tangential, radial, and line-of-sight directions for each chemically defined subpopulation. To sample the kinematics as a function of radius, stars in each subpopulation were divided into the same radial bins for all populations. The bin edges were defined based on the projected-radius distribution of the full sample using quantiles, so that the global sample is divided into bins containing comparable numbers of stars. The same fixed bin edges were then applied to each sub-population. The radial coordinate assigned to each bin corresponds to the midpoint of the lower and upper projected-radius boundaries of that bin. We adopted four radial bins for each profile as a compromise between spatial resolution and minimising low-number statistics; each bin typically contains $\sim$57–91 stars depending on the population. Although the parent APOGEE sample extends from $\sim0.9$ to $\sim40$ arcmin, the equal-number binning adopted here results in effective radial coverage between $\sim4$ and $\sim20$ arcmin ($\sim 4\,r_h$).

\begin{figure*}\centering
\includegraphics[clip=true, trim = 0mm 0mm 0mm 0mm, width=\linewidth]{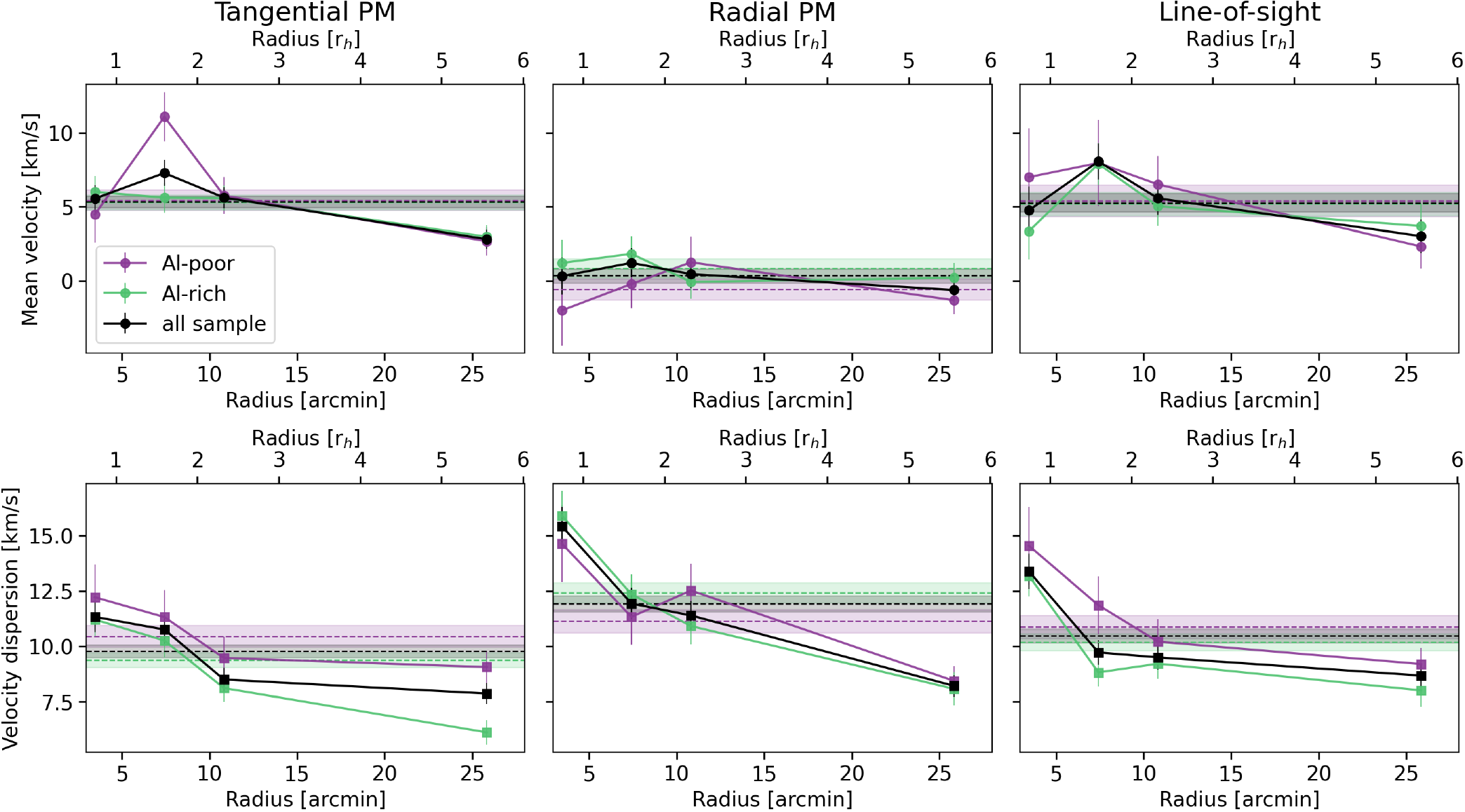}
  \caption{Kinematic profiles of the three velocity components. Top row: Mean tangential, radial proper motions, and line-of-sight velocity as a function of radius for Al-poor (purple), Al-rich (green), and all stars (black). Bottom row: Corresponding velocity dispersions. The shaded bands indicate the global values.}\label{fig:kinematic_profiles}
 \end{figure*}

For the proper-motion components, we modelled the velocity distribution in each radial bin as a bivariate Gaussian convolved with the individual measurement uncertainties, following the standard formalism adopted in Gaia-based dynamical analyses \citep[e.g.][]{vasiliev21}.  Posterior distributions of $v_{\mathrm{PM,rad}}$, $v_{\mathrm{PM,tan}}$, $\sigma_{\mathrm{PM,rad}}$, and $\sigma_{\mathrm{PM,tan}}$ were sampled using a Markov chain Monte Carlo (MCMC) approach with the \texttt{emcee} sampler \citep{foreman2013}, employing 32 walkers, 3000 steps, and a burn-in phase of 500 steps.  The line-of-sight velocity component ($v_{\mathrm{los}}$, $\sigma_{\mathrm{los}}$) was treated separately using a univariate Gaussian likelihood and the same inference procedure. The resulting radial profiles of the mean velocities and dispersions of the three velocity components for the Al-poor and Al-rich populations are shown in Fig.~\ref{fig:kinematic_profiles}.

\subsubsection{Global kinematics}
In addition to the radial profiles, we also computed global (radius-integrated) mean velocities and velocity dispersions for each subpopulation using the same modelling framework by considering all stars within a single radial bin.  These global values provide an integrated characterisation of the internal kinematics of each population and enable a direct comparison between the Al-poor and Al-rich groups.

\begin{figure*}\centering
\includegraphics[clip=true, trim = 0mm 0mm 0mm 0mm, width=0.47\linewidth]{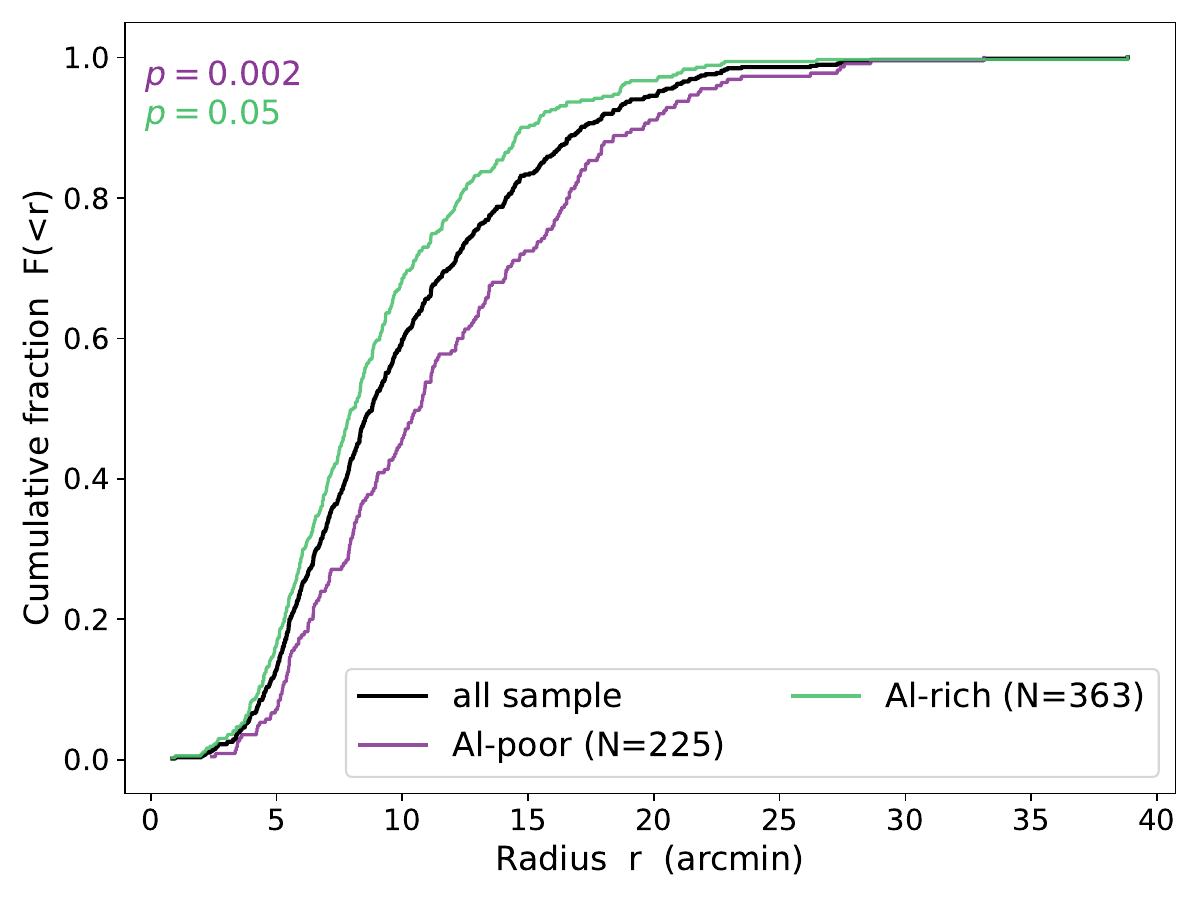}\\
\includegraphics[clip=true, trim = 0mm 0mm 0mm 0mm, width=0.33\linewidth]{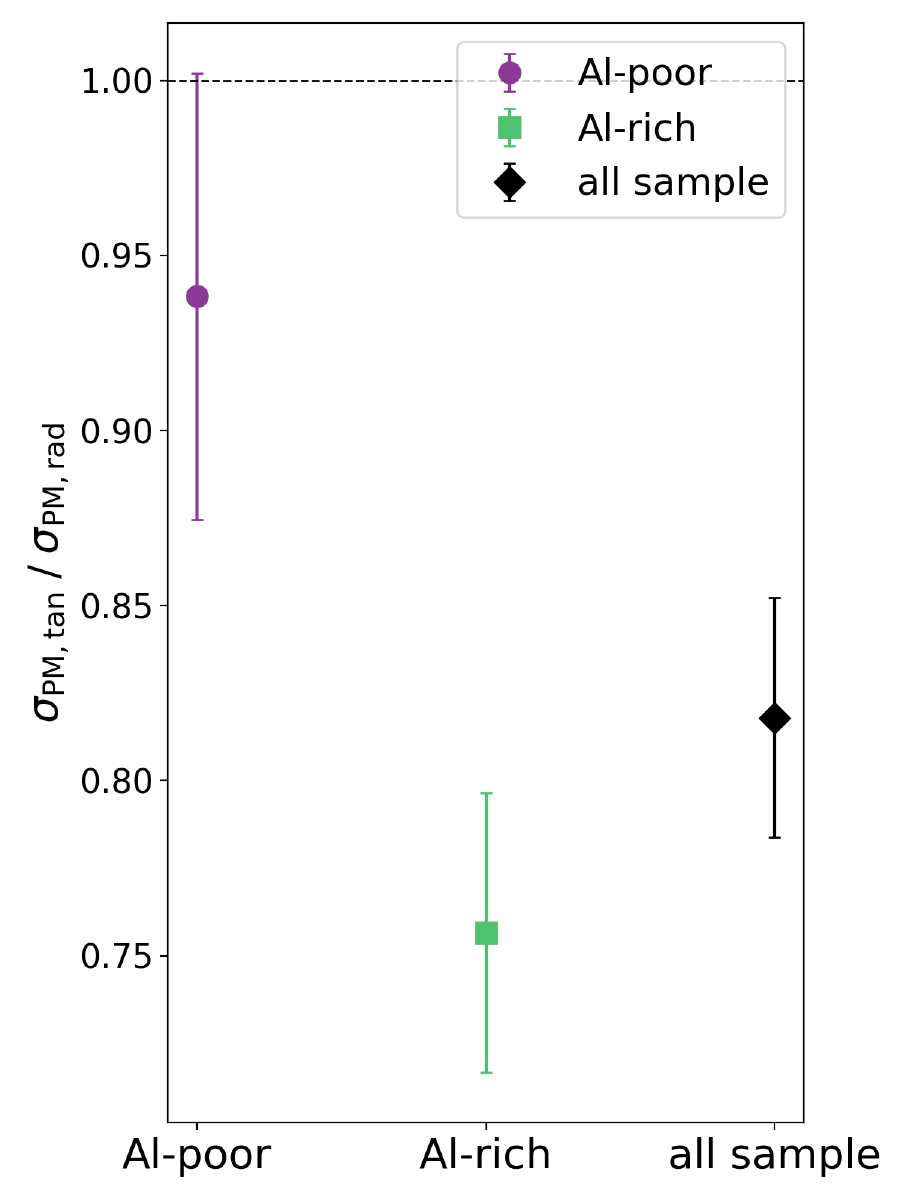}
\includegraphics[clip=true, trim = 0mm 0mm 0mm 0mm, width=0.33\linewidth]{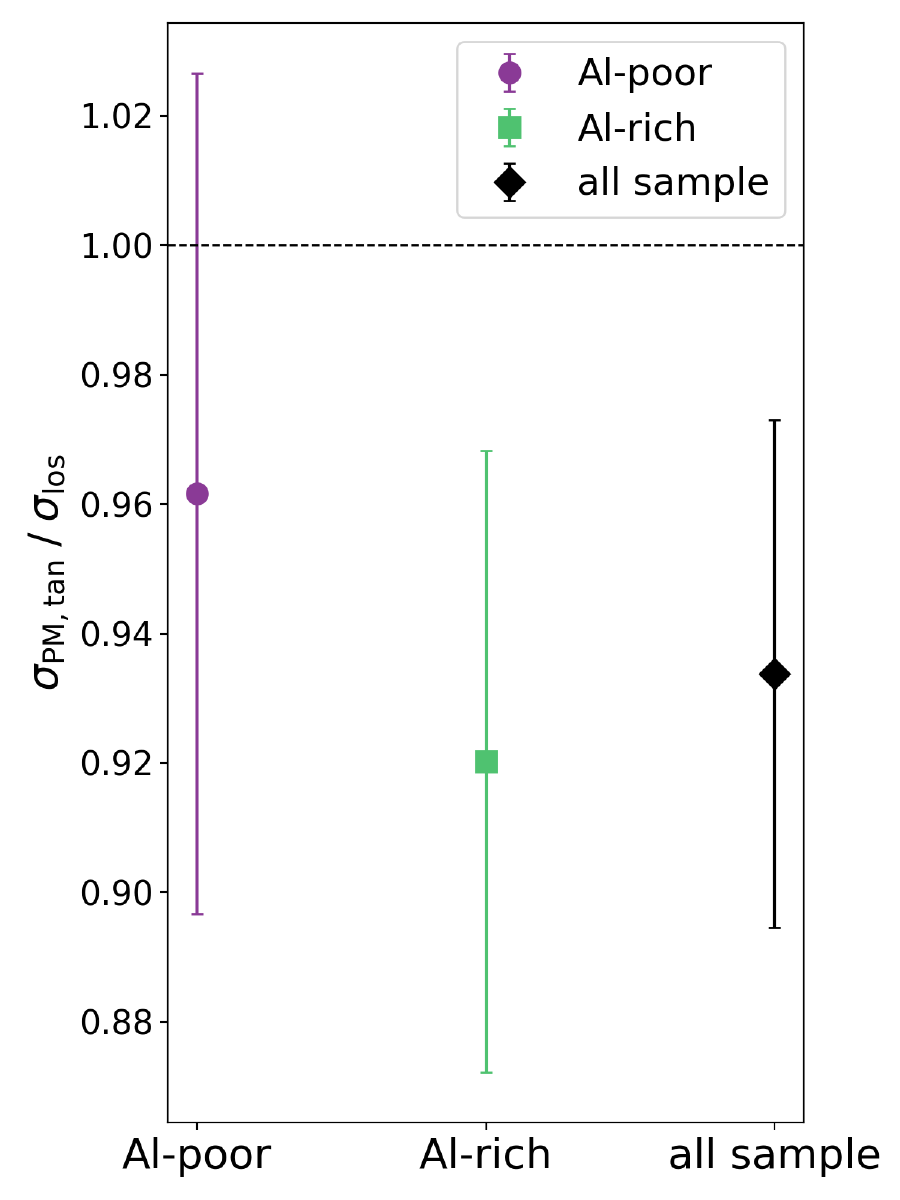}
\includegraphics[clip=true, trim = 0mm 0mm 0mm 0mm, width=0.33\linewidth]{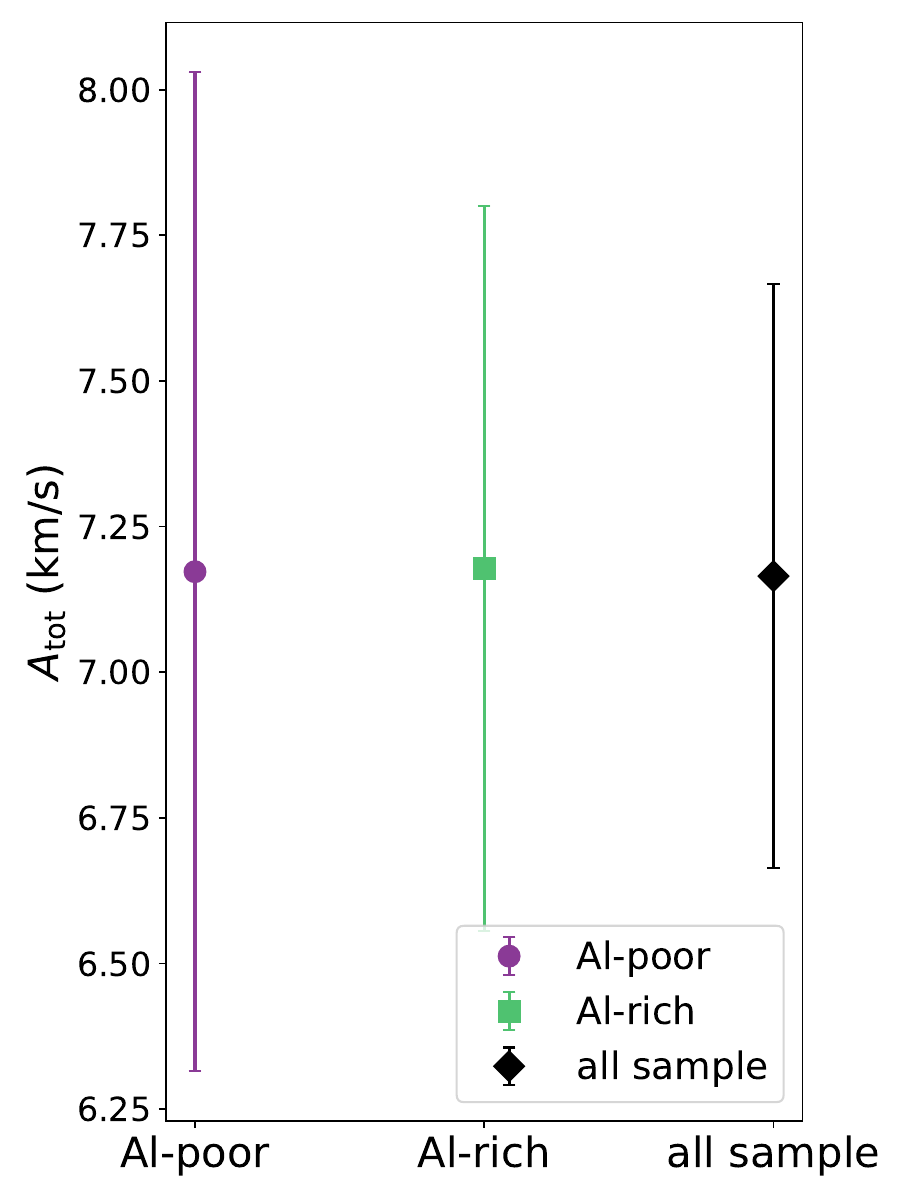}
  \caption{Spatial and global kinematic properties of the Al-poor and Al-rich populations. Top panel: Spatial cumulative distribution functions of the two families compared to the global sample. The p-values from a Kolmogorov-Smirnov test, which compares each family to the global sCDF, are also shown. Bottom panel: Global anisotropy ratios $\sigma_\mathrm{PM, tan}/\sigma_\mathrm{PM, rad}$, $\sigma_\mathrm{PM, tan}/\sigma_\mathrm{los}$ and total intrinsic rotation amplitude $A_\mathrm{tot}$ for the Al-poor, Al-rich, and combined samples. The figure shows that the Al-rich family is more centrally concentrated and more radially anisotropic than the Al-poor family, while both share a comparable intrinsic rotation amplitude.}\label{fig:2pops}
 \end{figure*}

\begin{figure*}
\sidecaption
  \includegraphics[width=12cm]{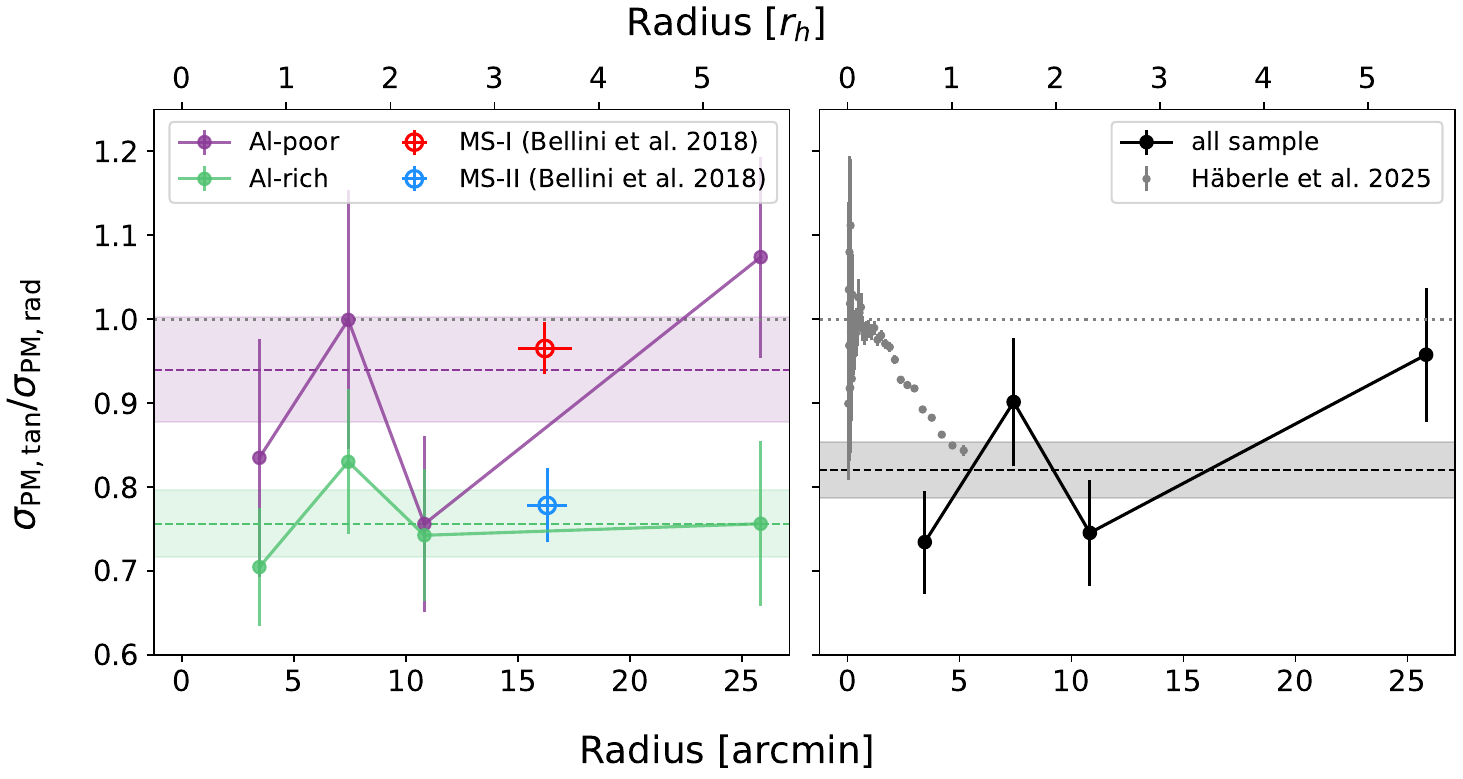}
     \caption{Radial anisotropy profiles of the two chemically defined populations. Left panel: $\sigma_{\mathrm{PM, tan}}/\sigma_{\mathrm{PM, rad}}$ as a function of radius for the Al-poor and Al-rich stars. The shaded bands mark the corresponding global values from bottom left panel of Fig.~\ref{fig:2pops}. The HST-based measurements of the MS-I and MS-II populations from \citet{Bellini2018} are overplotted for comparison. Right panel: Combined sample, compared to the oMEGACat profile from \citet{haberle2025}.}
     \label{fig:ani_profiles}
\end{figure*}

To assess differences in orbital structure, we quantified the velocity anisotropy of each subpopulation by computing ratios of the velocity dispersions in the tangential, radial, and line-of-sight directions. Specifically, we evaluated the ratios $\sigma_{\mathrm{PM,tan}}/\sigma_{\mathrm{PM,rad}}$ and $\sigma_{\mathrm{PM,tan}}/\sigma_{\mathrm{los}}$, which serve as proxies for the intrinsic orbital anisotropy in the available kinematic data. These ratios were measured both globally, using all stars in a single bin (see the bottom-left and central panels of Fig.~\ref{fig:2pops}), and as a function of radius by combining the dispersion profiles in multiple bins to construct anisotropy profiles (see Fig.~\ref{fig:ani_profiles}). While the global values provide a summary measurement of the internal kinematic structure, the radial profiles allow us to trace how orbital anisotropy varies with clustercentric distance.

\subsubsection{Reconstruction of the intrinsic rotation}
An alternative method to measure internal rotation, commonly used in the literature, is based on modelling the velocity field as a function of position angle on the sky. Following the approach of \citet{leitinger2025}, we modelled the line-of-sight velocity field with a sinusoidal function,
\begin{equation}
v_{\mathrm{los}}(\theta) = A_{\mathrm{LOS}} \, \sin(\theta - \theta_0),
\end{equation}
where $A_{\mathrm{LOS}}$ is the rotation amplitude projected along the line of sight and $\theta_0$ is the position angle of the rotation axis.

The tangential and radial proper-motion components were modelled as Gaussian distributions characterised by constant mean velocities and intrinsic dispersions. All three velocity components (radial PM, tangential PM, and LOS) were fitted simultaneously within a common MCMC framework implemented with the \texttt{emcee} sampler \citep{foreman2013}, using 32 walkers and 10\,000 steps, with the first 2\,000 steps discarded as burn-in. Posterior medians and standard deviations were adopted as best-fit values and uncertainties.

This procedure allows us to reconstruct the intrinsic 3D rotation by combining the LOS rotation amplitude $A_{\mathrm{LOS}}$ with the mean tangential velocity $\mu_{\mathrm{tan}}$, interpreted as orthogonal projections of the same rotation vector. The inclination of the rotation axis and the total intrinsic rotation amplitude were subsequently derived as
\begin{equation}
\tan i = \frac{A_{\mathrm{LOS}}}{\mu_{\mathrm{tan}}},
\qquad
A_{\mathrm{tot}} = \sqrt{A_{\mathrm{LOS}}^2 + \mu_{\mathrm{tan}}^2}.
\end{equation}
The resulting intrinsic rotation amplitudes for each chemically defined subpopulation are shown in the bottom-right panel of Fig.~\ref{fig:2pops}.

\begin{table*}[t]
\centering
\caption{Global anisotropy measurements, rotation amplitude, position angle of the rotation axis $\theta_0$, and inclination angle $i$ for the chemically
defined populations in $\omega$~Cen.}
\label{tab:anisotropy_global}
\begin{tabular}{lccccc}
\hline
Population & $\sigma_{\mathrm{PM, tan}}/\sigma_{\mathrm{PM, rad}}$ &  $\sigma_{\mathrm{PM, tan}}/\sigma_{\mathrm{los}}$ & A$_{\rm tot}$ [km/s]& $\theta_0\, [\degree]$& $i\, [\degree]$\\
\hline
Al-poor     & $0.94 \pm 0.06$ & $0.96 \pm 0.06$ & $7.17 \pm 0.86$ & $184 \pm 13$ & $41 \pm 7$\\
Al-rich     & $0.76 \pm 0.04$ & $0.92 \pm 0.05$ & $7.18 \pm 0.62$ & $162 \pm 9$ & $43 \pm 5$ \\
All sample  & $0.82 \pm 0.03$ & $0.94 \pm 0.04$ & $ 7.16 \pm 0.50$ & $171 \pm 8$ & $42 \pm 4$\\
\hline
\end{tabular}
\end{table*}

\begin{table}
\centering
\caption{Proper-motion anisotropy profiles ($\sigma_{\rm PM, tan}/\sigma_{\rm PM, rad}$), as shown in Fig~\ref{fig:ani_profiles}. }
\label{tab:pm_anisotropy_profiles_union}
\small
\begin{tabular}{lccc}
\toprule
$r$ [arcmin] & Al-poor & Al-rich & All sample \\
\midrule
3.45 & $0.83\pm0.14$ & $0.71\pm0.07$ & $0.74\pm0.06$ \\
7.42 & $1.01\pm0.16$ & $0.83\pm0.08$ & $0.90\pm0.08$ \\
10.82 & $0.75\pm0.11$ & $0.75\pm0.08$ & $0.74\pm0.06$ \\
25.84 & $1.07\pm0.12$ & $0.77\pm0.10$ & $0.96\pm0.08$ \\
\bottomrule
\end{tabular}
\end{table}

\section{Results}
\label{results}
The results presented in this work reveal a tight connection between chemical tagging and internal kinematics in $\omega$~Cen. Our APOGEE-based GMM identifies five components in multi-dimensional abundance space.  When we examine their [Al/Fe] and [Mg/Fe] distributions together with their radial cumulative profiles, these five components naturally group into two broader families, here labelled Al-poor and Al-rich (see Sect. \ref{analysis_1} and Fig. \ref{fig:5pops}).  The top-left panel of Fig.~\ref{fig:2pops} shows that the two groups occupy distinct regions of the [Al/Fe]-[Mg/Fe] plane, while the top-right panel illustrates that the Al-rich stars are more centrally concentrated than the Al-poor ones, which dominate at larger radii. A two-sample Kolmogorov-Smirnov (KS) test comparing each family to the global sCDF yields $p=0.002$ for the Al-poor population and $p=0.05$ for the Al-rich population, indicating statistically significant deviations from the global radial distribution. To quantify the statistical significance of the difference between the radial distributions of the Al-poor and Al-rich populations, we apply a two-sample Kolmogorov-Smirnov test to their cumulative radial distributions. The KS statistic measures the maximum vertical separation between the two cumulative distribution functions. The associated $p$-value quantifies the probability of obtaining a separation at least as large as the observed one under the null hypothesis that both populations are drawn from the same underlying radial distribution. For the two populations considered here, we obtain a $p$-value of $p = 4.23 \times 10^{-7}$, indicating that the observed difference between the cumulative radial distributions is very unlikely to arise from random sampling of a common parent distribution.  

Using APOGEE line-of-sight velocities combined with \emph{Gaia} proper motions, we derived global kinematic quantities for the Al-poor and Al-rich populations covering a radial range from $\sim$0.9 to $\sim$40 arcmin ($\sim 8\,r_h$). The bottom panels of Fig.~\ref{fig:2pops} report the resulting anisotropy ratios $\sigma_{\mathrm{PM, tan}}/\sigma_{\mathrm{PM, rad}}$, $\sigma_{\mathrm{PM, tan}}/\sigma_{\mathrm{los}}$, and the total intrinsic rotation amplitude $A_{\mathrm{tot}}$ (see also Table ~\ref{tab:anisotropy_global}). Both populations exhibit significant internal rotation and comparable total intrinsic rotation amplitudes within the uncertainties, implying that they are embedded in the same global angular-momentum structure. In contrast, their velocity anisotropy differs. The Al-rich population shows a systematically lower $\sigma_{\mathrm{PM, tan}}/\sigma_{\mathrm{PM, rad}}$ ratio than the Al-poor population, indicating a signature of radial anisotropy. The difference of anisotropy between the two populations corresponds to a significance of $2.43\,\sigma$, confirming a statistically meaningful separation between the two populations. This result is further supported by a permutation-based test, described in Appendix~\ref{stat_sig_pvalues}, which shows that the observed anisotropy contrast is unlikely to arise from a random partitioning of the data. When using the line-of-sight–based anisotropy proxy $\sigma_{\mathrm{PM, tan}}/\sigma_{\mathrm{los}}$, the difference between the two populations is much weaker, with a significance of only
$0.51\,\sigma$, indicating that the anisotropy contrast is not detected in the line-of-sight component\footnote{We also verified that the ratio $\sigma_{\mathrm{los}}/\sigma_{\mathrm{PM,rad}}$ shows the same qualitative population-dependent behaviour as $\sigma_{\mathrm{PM,tan}}/\sigma_{\mathrm{PM,rad}}$, with the Al-rich population exhibiting systematically lower values. This confirms that the anisotropy contrast does not depend on the specific projection used but reflects a robust difference in the orbital structure of the two populations.}. 

Figure~\ref{fig:ani_profiles} presents the radial behaviour of the anisotropy ratio $\sigma_{\mathrm{PM,tan}}/\sigma_{\mathrm{PM,rad}}$ for the two chemically defined populations. Across the full radial range probed, the Al-rich population consistently exhibits lower $\sigma_{\mathrm{PM,tan}}/\sigma_{\mathrm{PM,rad}}$ values than the Al-poor population, indicating a more radially biased orbital
distribution. In contrast, the Al-poor population remains consistent with an approximately isotropic velocity distribution at all radii. The shaded bands indicate the corresponding global anisotropy values, showing that the population-level separation seen in the global quantities is preserved when the kinematics are examined as a function of radius.

The kinematic profiles in Fig.~\ref{fig:kinematic_profiles} complement this population-resolved analysis. The mean tangential proper motion, radial proper motion, and line-of-sight velocity profiles of the Al-poor and Al-rich populations are consistent within the quoted uncertainties across the full radial range. Although individual radial bins show small deviations, we find no evidence for a coherent systematic offset between the population means in any of the three components, indicating that both families follow the same global rotation pattern.
The dispersion profiles reveal clearer population-dependent differences. The Al-rich stars exhibit systematically lower tangential velocity dispersion profiles and comparable, or slightly higher, radial dispersion profiles relative to the Al-poor population. This combination results in a stronger radial anisotropy for the Al-rich stars, while the Al-poor population maintains more similar dispersions in the radial and tangential components and therefore remains closer to isotropy. The line-of-sight dispersion profile follows a behaviour similar to that of the tangential component, with the Al-rich population tending to show slightly lower values, although globally the differences are smaller than in the proper-motion components.

We note that our kinematic analysis is based on projected quantities, which may limit the sensitivity to subtle kinematic differences. As a result, the observed differences should be regarded as lower limits. A full 3D analysis, including deprojection, is required to better characterise these effects. 
 
As an additional structural consistency check, we estimated the projected ellipticity of each population from the eigenvalues of the second-moment tensor of the stellar positions, with uncertainties derived via bootstrap resampling. We obtain $\epsilon = 0.10 \pm 0.05$ for the Al-poor stars and $\epsilon = 0.11 \pm 0.04$ for the Al-rich stars, with consistent major-axis position angles of ${\sim}4^\circ$. Within the uncertainties, the two populations therefore exhibit the same degree of flattening and the same projected orientation.

The kinematic analysis yields rotation-axis position angles of $\theta_0 = 184^\circ \pm 13^\circ$ (Al-poor) and $162^\circ \pm 9^\circ$ (Al-rich; see Table~\ref{tab:anisotropy_global}). The difference corresponds to $\sim1.4\sigma$ and is not statistically significant. The combined sample gives $\theta_0 = 171^\circ \pm 8^\circ$, consistent with both subpopulations. The inferred inclination angles are likewise similar ($i \simeq 41^\circ$–$43^\circ$; see Table~\ref{tab:anisotropy_global}). 

Taken together, the structural and kinematic measurements indicate that the Al-poor and Al-rich populations share the same projected geometry and 3D angular-momentum vector. We find no evidence for kinematically decoupled or misaligned components.
Recent work based on data from MUSE has suggested the presence of a counter-rotating core in $\omega$~Cen \citep{pechetti24}. The APOGEE sample analysed here does not probe the very central arcminutes of the cluster with sufficient spatial resolution to test this claim, and therefore our results neither confirm nor exclude the presence of such a feature.

\section{Discussion}
\label{discussion}
The spatial segregation between the two chemically defined families identified in this work is qualitatively consistent with previous studies based on HST ChMs and ground-based photometry, which show that chemically enriched or US populations are more centrally concentrated, while less enriched components become increasingly prominent at larger radii (e.g. \citealt{Bellini2018, cordoni2020, dondoglio2025}). Our APOGEE-based analysis extends this picture by demonstrating that a similar radial segregation is recovered using spectroscopically defined populations and traced over a much wider field of view. We note that the APOGEE DR16 sample analysed by \citet{meszaros2021} and discussed by \citet{alvarezgaray2024} shows limited spectroscopic coverage in the innermost region of $\omega$~Cen. In particular, the comparison presented in their Fig.~10 reveals a deficit of APOGEE targets within the central
arcminutes of the cluster. The APOGEE DR17 catalogue used in this work provides improved spatial sampling and includes stars down to projected radii of $\sim0.9$ arcmin from the cluster centre. Therefore, the enhanced central concentration of the Al-rich population observed in our analysis cannot be attributed to the observational bias discussed by \citet{alvarezgaray2024}, but instead reflects an intrinsic spatial segregation between the chemically defined populations.

Beyond this structural segregation, the anisotropy measurements provide a direct dynamical counterpart to the spatial differences. The comparison with the HST-based analysis of \citet{Bellini2018} is particularly instructive. Although their study is based on ChM selected MS stars and our analysis uses APOGEE red giants, the anisotropy levels are consistent within the uncertainties over the radial range where the datasets overlap (i.e. at $r \sim 3.5\,r_h$). In particular, in the left and central panels of Fig.~\ref{fig:ani_profiles}, we show that the anisotropy of our Al-poor population closely matches that of their MS-I component, which is identified as the less enriched, 1P-like population and is less centrally concentrated within the HST field. Conversely, the Al-rich population exhibits anisotropy values similar to those of their MS-II component, corresponding to a more chemically enriched (2P-like) population that is both more centrally concentrated and more radially biased. This trend is also consistent with \citet{cordoni2020}, who find that light-element-senriched (N-rich) stars are more radially anisotropic than the less enriched (N-poor) population. The recovery of the same population-dependent anisotropy pattern across different stellar evolutionary stages and independent datasets strengthens the connection between the Al-poor and Al-rich division adopted here and the multiple populations traced by the ChM in $\omega$~Cen.

The right panel of Fig.~\ref{fig:ani_profiles} provides an additional consistency check by comparing our results with the oMEGACat kinematics of \citet{haberle2025}. Their combined MUSE and HST measurements show that $\omega$~Cen is approximately isotropic in the core and becomes increasingly radially anisotropic towards the half-light radius. Our combined APOGEE and Gaia profile follows the same behaviour where the datasets overlap, but the population-resolved view reveals that this global trend is not uniform: the Al-rich group carries most of the increasing radial anisotropy, while the Al-poor stars remain close to isotropic. Because the oMEGACat data primarily probe the inner few central arcminutes, whereas our combined APOGEE and Gaia sample extends from the crowding-limited core to $~\sim 4\, r_h $, we obtain a nearly continuous view of the radial behaviour of the two populations. 

One of the most striking chemical properties of $\omega$~Cen is its large internal iron spread, spanning more than one dex in [Fe/H]. Previous studies have investigated the relation between rotation and metallicity in $\omega$~Cen, finding no strong systematic dependence of the rotation amplitude on [Fe/H] \citep{pancino07}. Consistent with this picture, we do not detect significant differences in the rotation properties when subdividing the sample either by iron abundance or by aluminium enrichment. Instead, the population-dependent kinematic differences identified in this work are primarily seen in the velocity anisotropy. When dividing the sample by metallicity alone, we did not detect statistically significant differences in anisotropy. In contrast, a clear anisotropy difference emerges when the populations are separated according to aluminium enrichment, with the Al-rich stars exhibiting more radially biased orbits than the Al-poor population \citep[see also][]{cordoni2020, haberle2025}. This behaviour contrasts with that observed in systems such as M54, where kinematic differences are more closely linked to metallicity \citep[e.g.][]{alfaro-cuello2019}, suggesting that similar present-day nuclear clusters may arise from different chemo-dynamical formation pathways. 

Furthermore, the fact that population-dependent kinematic signatures are still observed today suggests that these structures are long-lived. In particular, the systematic difference in velocity anisotropy between Al-poor and Al-rich populations, recovered over a wide radial range, indicates that the internal dynamical evolution of $\omega$~Cen has not completely erased the imprint of population-specific orbital properties. This behaviour is also qualitatively consistent with theoretical models of the long-term dynamical evolution of MPs in GCs. Numerical simulations \citep[see e.g.][]{mastrobuono2013, tiongco2019, mastrobuono2021, vesperini2021} show that when one population is initially more centrally concentrated than another, the two components naturally evolve towards different velocity-anisotropy profiles: the concentrated population develops a radially anisotropic velocity distribution, while the more extended population remains closer to isotropic or mildly tangentially biased. Importantly, these anisotropy differences are predicted to be among the most long-lived kinematic signatures of incomplete dynamical mixing and can persist over several relaxation times \citep{aros2025}. The spatial and anisotropy ordering observed here between the Al-rich and Al-poor populations therefore follows the qualitative expectations of such dynamical evolution models.

The coupled chemical, spatial, and kinematic properties revealed by our analysis provide important constraints on the formation history of $\omega$~Cen. In particular, the APOGEE-defined Al-poor and Al-rich populations differ in their spatial distributions and orbital anisotropies, while simultaneously sharing a common large-scale rotation pattern. This combination of similarities and differences offers insight into the physical processes that shaped the system.

One possibility is that $\omega$~Cen formed through the hierarchical assembly of multiple stellar systems inside a larger host galaxy (e.g. \citealt{bekki03, bekki2016, calamida2020}). In this picture, inspiral and merging can generate spatial segregation and incomplete dynamical mixing, leaving residual differences in orbital structure. The stronger central concentration and radial anisotropy of the Al-rich population are qualitatively consistent with such a process. However, if mergers were the dominant driver, one might expect the clearest chemo-dynamical differences to correlate with metallicity, whereas in our data they are more strongly linked to aluminium enrichment. This suggests that hierarchical assembly, if contributed, is unlikely to explain the observed chemo-dynamical structure on its own.

More generally, mergers alone do not straightforwardly account for the large metallicity spread and the extreme light-element enrichment observed in $\omega$~Cen, both of which point to efficient gas retention and prolonged chemical evolution. In particular, the large Al spread is closely connected to the multiple-population phenomenon in GCs and is generally interpreted as the result of internal self-enrichment. We also note that other massive GCs not suspected to be the nuclei of stripped dwarf galaxies exhibit comparable Al variations (e.g. NGC~2808, NGC~6752, and NGC~6205). A more detailed chemo-dynamical characterisation of such systems may therefore help place the properties of $\omega$~Cen in a broader context.

A complementary, and not mutually exclusive, interpretation is that $\omega$~Cen retained gas over an extended period and experienced multiple episodes of centrally concentrated star formation (e.g. \citealt{romano2007,marcolini2007,gratton2012}). In this view, the Al-poor stars would correspond to an earlier generation with weaker light-element anomalies, while the Al-rich stars would have formed later from enriched gas. This scenario naturally explains the stronger central concentration of the Al-rich population and is also consistent with its distinct orbital anisotropy, while the shared large-scale rotation pattern suggests that both populations evolved within the same global potential. The chemical-evolution models of \citet{dondoglio2025} provide a quantitative framework consistent with this interpretation.

Taken together, the observational constraints presented here could be consistent with a hybrid formation pathway in which both hierarchical assembly and prolonged internal star formation contributed to the build-up of $\omega$~Cen. The shared rotation pattern strongly suggests that the chemically distinct populations evolved within a common gravitational potential, while their differing spatial distributions and orbital anisotropies indicate that their dynamical histories were not identical.

The correspondence between our APOGEE-defined populations and the chemical groups identified by \citet{Mason2025} provides additional insight into the formation sequence of $\omega$~Cen. In their framework, the P1 population represents the least chemically processed component, while the IM and P2 populations correspond to progressively more enriched stellar generations. Our Al-poor stars broadly map onto their P1 population, whereas our Al-rich sample appears to encompass both their IM and P2 components. Within this interpretation, the observed chemo-dynamical structure may be naturally explained by a multi-phase formation history. A first major episode of star formation could have produced the P1 or Al-poor population, which today appears more spatially extended and closer to isotropic. Subsequent evolution may then have involved either (i) the accretion of relatively pristine gas that triggered further centrally concentrated star formation, producing chemically enriched stars with higher aluminium abundances, or (ii) the inward migration and merging of additional stellar clusters with IM-like chemistry, which could both deepen the central potential and contribute gas reservoirs that fuel the formation of the most chemically extreme P2 population.

In either case, the IM and P2 components -- corresponding to our Al-rich population -- would be expected to form or settle preferentially in the central regions and to develop partially distinct orbital properties with respect to the P1 or Al-poor population. The persistence of their stronger radial anisotropy may therefore reflect either their later formation within a deeper central potential or their dynamical deposition through inspiral processes.

The origin of the most metal-rich and Al-enhanced populations remains an open question. A merger origin for these components would require the accreted system to share a very similar angular-momentum structure with the pre-existing cluster, in order to reproduce the observed common rotation axis and inclination. Such a configuration is not impossible, but it would imply a degree of dynamical alignment that may require fine tuning. Conversely, a scenario in which metal-rich and Al-enhanced stars formed in situ from centrally retained and progressively enriched gas provides a more straightforward explanation for their shared rotation geometry and enhanced central concentration.

\section{Conclusions}

\label{conclusions}
This work provides the first wide-field, fully spectroscopic view of population-dependent 3D internal kinematics of $\omega$~Cen.
We combined APOGEE DR17 abundances with Gaia kinematics to study the chemo-dynamical structure of $\omega$~Cen. A GMM applied to 8D abundance space identifies five components, which, after examining their radial cumulative distributions, we grouped into two broader families that we refer to as Al-poor and Al-rich. These chemically defined families also differ in their spatial and kinematic properties.

Our main findings are as follows:
\begin{enumerate}
    \item The Al-rich stars are significantly more centrally
    concentrated than the Al-poor stars, which dominate at larger radii.

    \item The Al-rich population exhibits stronger radial anisotropy and
largely drives the global anisotropy trend of $\omega$~Cen, whereas the
Al-poor stars remain closer to isotropic across the radial range probed.
    \item The chemo-dynamical differences correlate primarily with
    light-element enrichment rather than with iron abundance.
    Despite the well-known internal [Fe/H] spread in $\omega$~Cen,
    the spatial segregation and anisotropy differences identified here
    are more associated with aluminium enrichment than with
    metallicity itself, indicating that light-element chemistry is the
    dominant tracer of the dynamical substructure.
    \item Despite their different anisotropies, the two populations share
    a common rotation pattern, suggesting that they are embedded in the
    same global angular-momentum structure. Their total rotation amplitudes are
    comparable within the uncertainties, and we find no evidence for kinematically decoupled or counter-rotating
    components in our large field of view. 
\end{enumerate}

Taken together, these chemo-dynamical constraints are difficult to reconcile with a simple, monolithic globular-cluster origin. 
Instead, they support a formation pathway in which $\omega$~Cen evolved within a deep potential well capable of sustaining prolonged chemical enrichment and internal dynamical differentiation. 
In this context, our results are compatible with self-enrichment models proposed for $\omega$~Cen \citep{dondoglio2025}, in which extended star formation within a dwarf-galaxy environment gives rise to multiple chemically distinct but dynamically coupled populations. 
A sufficiently deep progenitor potential could retain a fraction of supernova ejecta, enabling progressive enrichment and naturally accounting for the observed internal [Fe/H] spread. 
Within such a framework, the Al-poor stars likely trace an earlier, more spatially extended component, whereas the Al-rich population formed later from centrally concentrated, chemically processed gas. 
The shared rotation geometry and consistent angular-momentum orientation indicate that both populations evolved within the same global gravitational potential, while their distinct orbital anisotropies reflect incomplete dynamical mixing rather than independent dynamical origins.

Future work should combine extended spectroscopic samples, especially in the outermost regions of the cluster, with orbit-based or $N$-body models that explicitly treat multiple tracer populations. Such chemo-dynamical modelling will be essential to quantify the relative roles of mergers and internal chemical evolution in assembling $\omega$~Cen, and to place this system in the broader context of NSC formation in dwarf galaxies.

Finally, we note that our work provides a fully spectroscopic way to characterise the chemo-dynamical properties of GCs, independent on the widely used photometric detections of MPs. An application of our method to a wider samples of MW GCs could further shed light onto the puzzling formation of these dense stellar systems.


\begin{acknowledgements}
G.P. acknowledges the support from the Centre national d'études spatiales (CNES) through a postdoctoral fellowship. G.P., P.B., and P.D.M. are grateful to the ``Action Thématique de Cosmologie et Galaxies'' (ATCG), Programme National ASTRO of the INSU (Institut National des Sciences de l’Univers) for supporting this research, in the framework of the project "Coevolution of globular clusters and dwarf galaxies, in the context of hierarchical galaxy formation: from the Milky Way to the nearby Universe", PI: A. Lançon (2025), G. Pagnini (2026). P.B. and G.P. acknowledge financial support by the IdEx framework of the University of Strasbourg. Funding for the Sloan Digital Sky Survey IV has been provided by the Alfred P. Sloan Foundation, the U.S. Department of Energy Office of Science, and the Participating Institutions. SDSS-IV acknowledges support and resources from the Center for High Performance Computing at the University of Utah. The SDSS website is \href{}{www.sdss.org}. SDSS-IV is managed by the Astrophysical Research Consortium for the Participating Institutions of the SDSS Collaboration including the Brazilian Participation Group, the Carnegie Institution for Science, Carnegie Mellon University, Center for Astrophysics | Harvard \& Smithsonian, the Chilean Participation Group, the French Participation Group, Instituto de Astrof\'isica de Canarias, The Johns Hopkins University, Kavli Institute for the Physics and Mathematics of the Universe (IPMU) / University of Tokyo, the Korean Participation Group, Lawrence Berkeley National Laboratory, Leibniz Institut f\''ur Astrophysik Potsdam (AIP),  Max-Planck-Institut f\''ur Astronomie (MPIA Heidelberg), Max-Planck-Institut f\''ur Astrophysik (MPA Garching), Max-Planck-Institut f\''ur Extraterrestrische Physik (MPE), National Astronomical Observatories of China, New Mexico State University, New York University, University of Notre Dame, Observat\'ario Nacional / MCTI, The Ohio State University, Pennsylvania State University, Shanghai Astronomical Observatory, United Kingdom Participation Group, Universidad Nacional Aut\'onoma de M\'exico, University of Arizona, University of Colorado Boulder, University of Oxford, University of Portsmouth, University of Utah, University of Virginia, University of Washington, University of Wisconsin, Vanderbilt University, and Yale University. This work has made use of data from the European Space Agency (ESA) mission \textit{Gaia} (\href{}{https://www.cosmos.esa.int/gaia}), processed by the \textit{Gaia} Data Processing and Analysis Consortium (DPAC, \href{}{https://www.cosmos.esa.int/web/gaia/dpac/consortium}). Funding for the DPAC has been provided by national institutions, in particular the institutions participating in the \textit{Gaia} Multilateral Agreement. 
\end{acknowledgements}

\bibliographystyle{aa}
\bibliography{biblio}

\clearpage

 
 
 
 \begin{appendix}
\section{8D chemical distribution of $\omega$~Cen and its subpopulations}
\begin{figure}\centering
\includegraphics[clip=true, trim = 0mm 0mm 0mm 0mm, width=\linewidth]{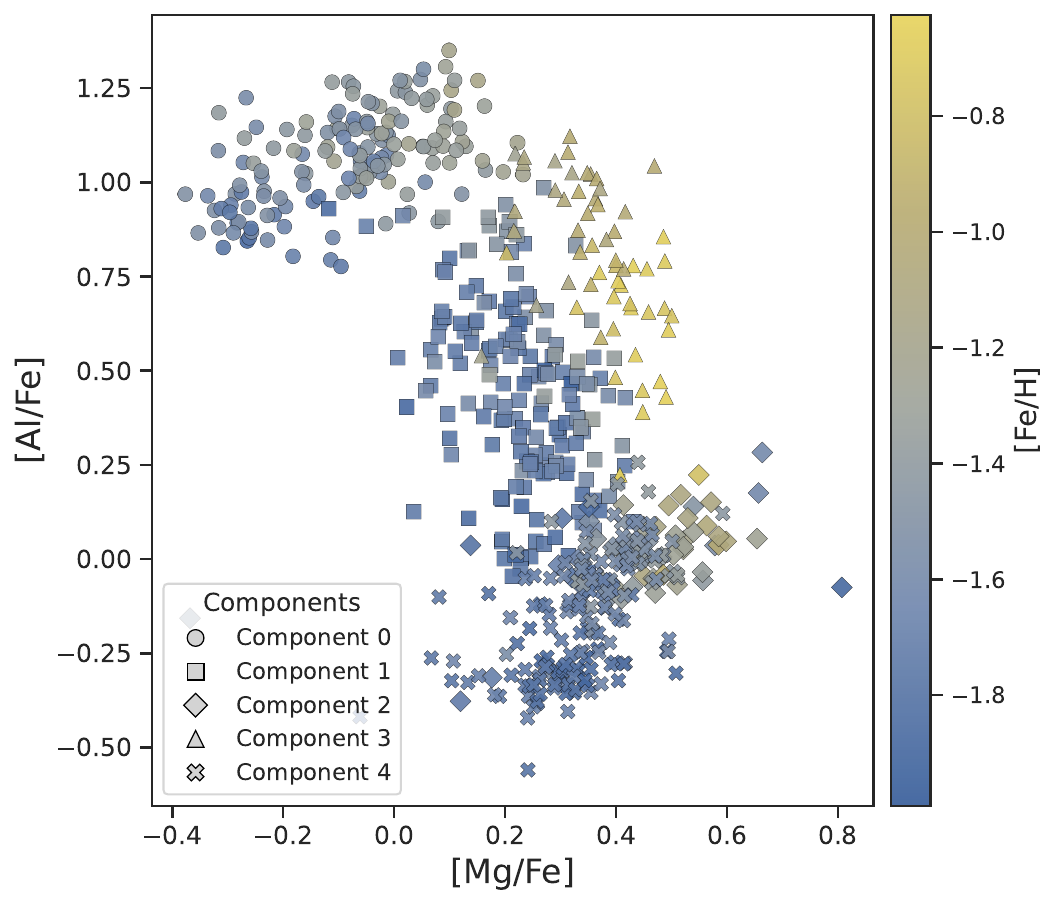}
  \caption{Alternative representation of the left panel of Fig.~\ref{fig:5pops} in the [Mg/Fe]–[Al/Fe] plane. Stars are shown with different marker symbols according to their GMM chemical component, while the colour scale encodes metallicity, [Fe/H].}\label{fig:mgfe_alfe_feh}
 \end{figure}

\begin{figure*}\centering
\includegraphics[clip=true, trim = 0mm 0mm 0mm 0mm, width=\linewidth]{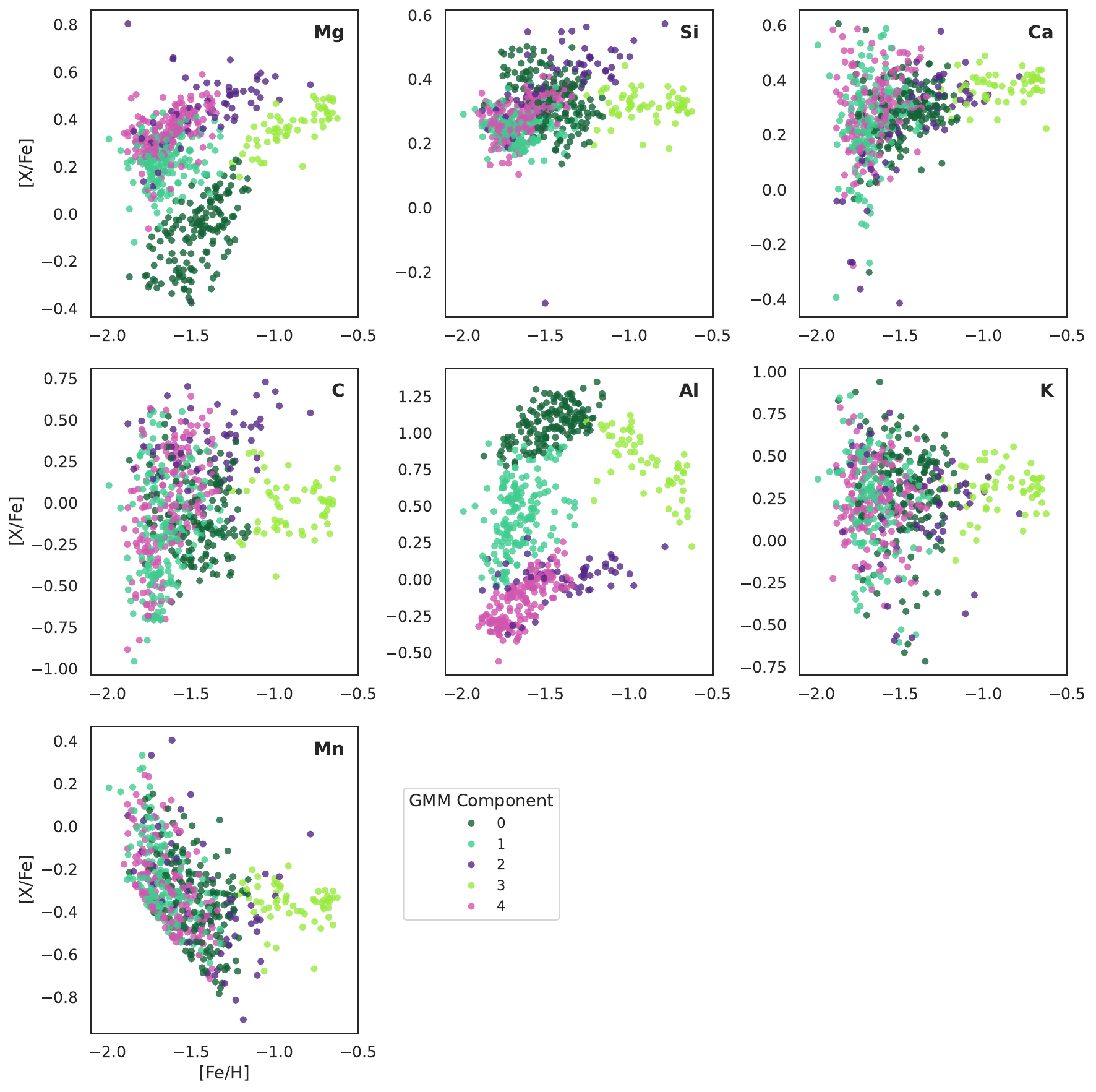}
  \caption{Projections of the 8D chemical-abundance space used for the GMM analysis of $\omega$~Cen APOGEE stars. Each panel shows $[\mathrm{X/Fe}]$ as a function of $[\mathrm{Fe/H}]$ for one of the elements included in the clustering (Mg, Si, Ca, C, Al, K, and Mn). The points are colour-coded according to the five GMM components. }\label{fig:8dchem}
 \end{figure*}

Appendix~A summarises the multi-dimensional chemical distribution of the APOGEE $\omega$~Cen sample analysed in this work. Figure~\ref{fig:mgfe_alfe_feh} shows an alternative representation of the $[\mathrm{Mg/Fe}]$--$[\mathrm{Al/Fe}]$ plane, in which different marker symbols identify the GMM chemical components and the colour scale traces $[\mathrm{Fe/H}]$, thereby highlighting the internal metallicity spread within each population. Figure~\ref{fig:8dchem} shows the projections of the 8D abundance space onto the individual $[\mathrm{X/Fe}]$--$[\mathrm{Fe/H}]$ planes for all analysed elements, colour-coded according to the Gaussian-mixture components, and illustrates how the five chemically tagged populations occupy distinct regions of abundance space. Table~\ref{tab:apogee_ids} lists the APOGEE identifiers together with the corresponding GMM component labels for the full sample.

 \begin{table}[h!]\centering
\begin{tabular}{lc}
\toprule
APOGEE ID & component \\
\midrule
2M13233428-4734432 & 4 \\
2M13240738-4724118 & 4 \\
2M13242041-4719352 & 1 \\
2M13243074-4724264 & 4 \\
2M13243893-4731381 & 4 \\
2M13244170-4728147 & 4 \\
2M13244375-4721582 & 4 \\
2M13244413-4719252 & 1 \\
2M13244491-4725264 & 4 \\
2M13244695-4724580 & 4 \\
..... & ......\\
\bottomrule
\end{tabular}
\caption{APOGEE IDs and GMM labels for $\omega$ Centauri sample. This table is available in its entirety at the CDS.}\label{tab:apogee_ids}
\end{table}
 
 \section{Statistical significance of population-dependent spatial and kinematic differences}\label{stat_sig_pvalues}

To assess the statistical significance of the difference in velocity anisotropy between the Al-poor and Al-rich populations, we adopt a permutation-based approach.

We define a proxy anisotropy parameter as
\[
\beta^\prime \equiv \sigma_{\mathrm{PM, tan}} / \sigma_{\mathrm{PM, rad}},
\]
where $\sigma_{\mathrm{tan}}$ and $\sigma_{\mathrm{rad}}$ represent the tangential and radial PM component, respectively. We quantify the anisotropy contrast between the two populations as
\[
\Delta \beta^\prime =
\left( \beta^\prime \right)_{\mathrm{Al\text{-}poor}}
-
\left( \beta^\prime \right)_{\mathrm{Al\text{-}rich}}.
\]

To construct a null distribution for $\Delta \beta^\prime$, stars are randomly reassigned between two populations 100 times while keeping the population sizes fixed. For each random realisation, we recompute $\Delta \beta^\prime$, thereby building the distribution of anisotropy contrasts expected under the null hypothesis that population membership is unrelated to orbital anisotropy.

\begin{figure}\centering
\includegraphics[clip=true, trim = 0mm 0mm 0mm 0mm, width=\linewidth]{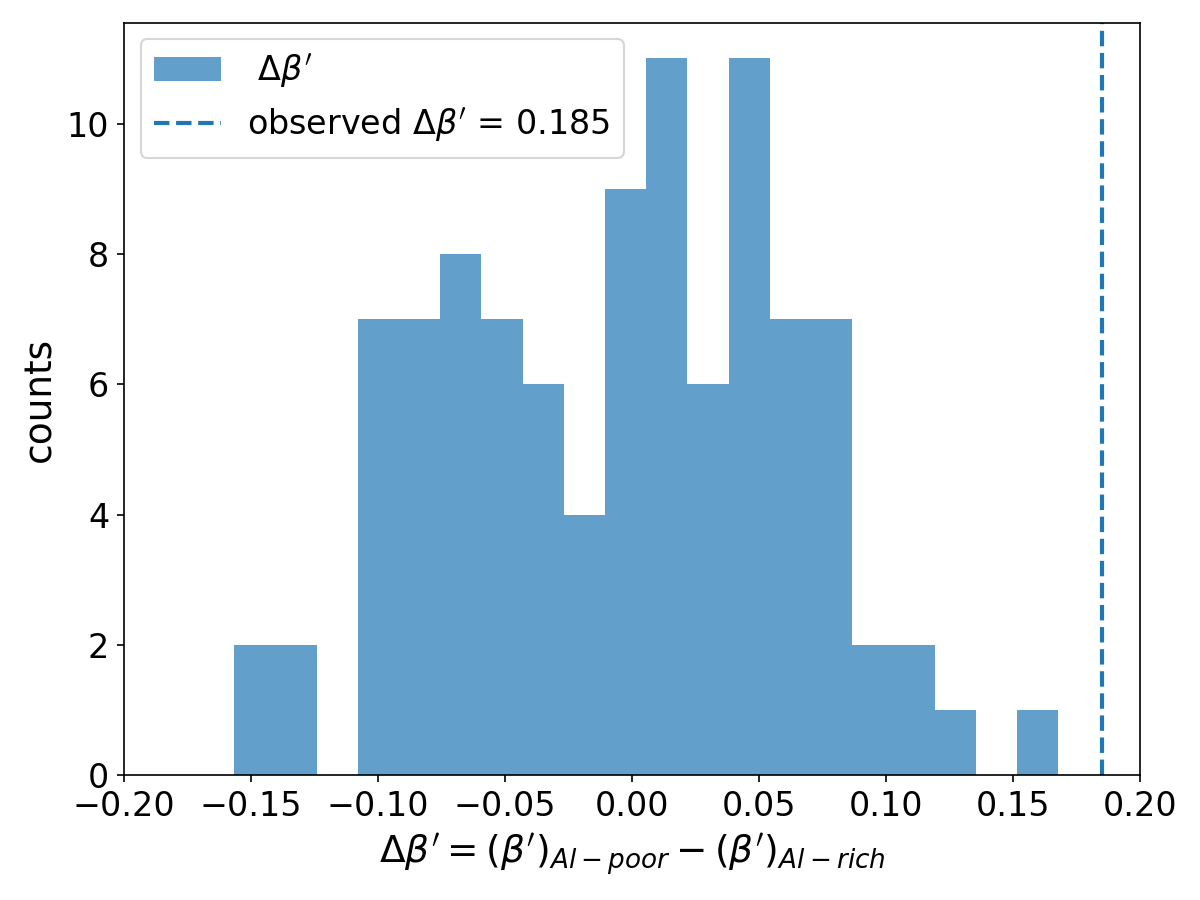}
  \caption{Distribution of anisotropy contrasts. The $\Delta \beta^\prime$ values are obtained from the random reshuffling of stars between the Al-poor and Al-rich populations while keeping population sizes fixed.
The vertical dashed line marks the observed anisotropy contrast.}\label{fig:delta_anisotropy}
 \end{figure}

Figure~\ref{fig:delta_anisotropy} shows the resulting null distribution of $\Delta \beta^\prime$. The observed anisotropy contrast, $\Delta \beta^\prime \simeq 0.19$, is indicated by a vertical dashed line and lies well outside the bulk of the distribution obtained from random reshuffling. The corresponding $p$-value is defined as the fraction of random realisations that yield a value of $|\Delta \beta^\prime|$ at least as large as the observed one.

This result indicates that the measured difference in velocity anisotropy between the Al-poor and Al-rich populations is unlikely to arise from a random partitioning of the data, supporting the interpretation that the two populations occupy systematically different regions of orbital phase space.

\section{Kinematic profiles}
Appendix~C provides the numerical values underlying the kinematic profiles shown in Fig.~\ref{fig:kinematic_profiles}. Tables~\ref{tab:kin_profiles_means} and \ref{tab:kin_profiles_sigmas} report the mean velocities and velocity dispersions, respectively, for the tangential proper-motion, radial proper-motion, and line-of-sight components, both globally and in each radial bin for the Al-poor, Al-rich, and combined samples. 

\begin{table*}
\centering
\caption{Mean velocity profiles and global values for the three velocity components, as shown in the first row of Fig.~\ref{fig:kinematic_profiles}.}
\label{tab:ocen_vel_profiles_wide}
\small
\setlength{\tabcolsep}{3.5pt}
\renewcommand{\arraystretch}{1.05}
\begin{tabular}{lccc ccc ccc}
\toprule
& \multicolumn{3}{c}{All sample [km/s]} & \multicolumn{3}{c}{Al-poor [km/s]} & \multicolumn{3}{c}{Al-rich [km/s]} \\
\cmidrule(lr){2-4}\cmidrule(lr){5-7}\cmidrule(lr){8-10}
$r$ [arcmin] & $v_{\rm PM,tan}$ & $v_{\rm PM,rad}$ & $v_{\rm los}$ & $v_{\rm PM,tan}$ & $v_{\rm PM,rad}$ & $v_{\rm los}$ & $v_{\rm PM,tan}$ & $v_{\rm PM,rad}$ & $v_{\rm los}$ \\
\midrule
\textit{global} & $5.33\pm0.39$ & $0.33\pm0.48$ & $5.27\pm0.63$ & $5.45\pm0.70$ & $-0.58\pm0.72$ & $5.41\pm1.07$ & $5.29\pm0.50$ & $0.85\pm0.65$ & $5.20\pm0.78$ \\
3.45 & $5.56\pm0.92$ & $0.31\pm1.24$ & $4.76\pm1.61$ & $4.48\pm1.87$ & $-2.02\pm2.38$ & $7.01\pm3.34$ & $6.02\pm1.08$ & $1.20\pm1.56$ & $3.35\pm1.90$ \\
7.42 & $7.29\pm0.89$ & $1.21\pm0.98$ & $8.07\pm1.23$ & $11.10\pm1.67$ & $-0.23\pm1.64$ & $7.97\pm2.93$ & $5.61\pm1.02$ & $1.82\pm1.22$ & $7.91\pm1.42$ \\
10.82 & $5.62\pm0.70$ & $0.44\pm0.94$ & $5.57\pm1.11$ & $5.76\pm1.25$ & $1.24\pm1.74$ & $6.52\pm1.93$ & $5.59\pm0.84$ & $-0.08\pm1.13$ & $5.04\pm1.36$ \\
25.84 & $2.81\pm0.66$ & $-0.64\pm0.68$ & $3.02\pm1.12$ & $2.66\pm0.98$ & $-1.33\pm0.93$ & $2.32\pm1.52$ & $2.97\pm0.78$ & $0.19\pm1.03$ & $3.70\pm1.53$ \\
\bottomrule
\end{tabular}\label{tab:kin_profiles_means}
\end{table*}

\begin{table*}
\centering
\caption{Dispersion profiles and global values for the three components as shown in the second row of Fig.~\ref{fig:kinematic_profiles}.}
\label{tab:ocen_disp_profiles_wide}
\small
\setlength{\tabcolsep}{3.5pt}
\renewcommand{\arraystretch}{1.05}
\begin{tabular}{lccc ccc ccc}
\toprule
& \multicolumn{3}{c}{All sample [km/s]} & \multicolumn{3}{c}{Al-poor [km/s]} & \multicolumn{3}{c}{Al-rich [km/s]} \\
\cmidrule(lr){2-4}\cmidrule(lr){5-7}\cmidrule(lr){8-10}
$r$ [arcmin] & $\sigma_{\rm PM,tan}$ & $\sigma_{\rm PM,rad}$ & $\sigma_{\rm los}$ & $\sigma_{\rm PM,tan}$ & $\sigma_{\rm PM,rad}$ & $\sigma_{\rm los}$ & $\sigma_{\rm PM,tan}$ & $\sigma_{\rm PM,rad}$ & $\sigma_{\rm los}$ \\
\midrule
\textit{global} & $9.78\pm0.29$ & $11.93\pm0.35$ & $10.47\pm0.30$ & $10.46\pm0.50$ & $11.14\pm0.54$ & $10.88\pm0.52$ & $9.37\pm0.34$ & $12.43\pm0.46$ & $10.19\pm0.38$ \\
3.45 & $11.34\pm0.68$ & $15.44\pm0.89$ & $13.40\pm0.80$ & $12.23\pm1.47$ & $14.65\pm1.75$ & $14.55\pm1.76$ & $11.21\pm0.79$ & $15.92\pm1.11$ & $13.20\pm0.93$ \\
7.42 & $10.76\pm0.65$ & $11.94\pm0.71$ & $9.72\pm0.57$ & $11.33\pm1.22$ & $11.34\pm1.26$ & $11.86\pm1.31$ & $10.26\pm0.77$ & $12.36\pm0.90$ & $8.82\pm0.64$ \\
10.82 & $8.49\pm0.52$ & $11.40\pm0.66$ & $9.50\pm0.55$ & $9.47\pm0.95$ & $12.53\pm1.21$ & $10.21\pm1.03$ & $8.11\pm0.61$ & $10.92\pm0.82$ & $9.21\pm0.70$ \\
25.84 & $7.86\pm0.46$ & $8.21\pm0.49$ & $8.67\pm0.52$ & $9.06\pm0.72$ & $8.43\pm0.66$ & $9.19\pm0.73$ & $6.10\pm0.56$ & $8.07\pm0.75$ & $8.00\pm0.74$ \\
\bottomrule
\end{tabular}\label{tab:kin_profiles_sigmas}
\end{table*}

\end{appendix}

\end{document}